\documentclass{emulateapj}

\usepackage{amssymb}
\usepackage{amsmath}
\usepackage{graphicx}
\usepackage{natbib}
\usepackage{txfonts}
\usepackage[colorlinks=true,citecolor=blue,linkcolor=blue]{hyperref}
\usepackage{microtype}
\usepackage{color}

\def\mnras{MNRAS}
\def\apj{ApJ}
\def\apjl{ApJL}

\def\aap{A\&A}

\begin{document}

\title{On the Induced Gravitational Collapse of a Neutron Star to a Black Hole by a Type Ib/c Supernova}
\author{Jorge A.~Rueda and Remo Ruffini}
\affil{Dipartimento di Fisica and ICRA, Sapienza Universit\`a di Roma, P.le Aldo Moro 5, I–-00185 
Rome, Italy\\ICRANet, P.zza della Repubblica 10, I--65122 Pescara, Italy}
\altaffiltext{}{jorge.rueda@icra.it, ruffini@icra.it}

\begin{abstract}
It is understood that the Supernovae (SNe) associated to Gamma Ray Bursts (GRBs) are of type Ib/c. The temporal coincidence of the GRB and the SN represents still a major enigma of Relativistic Astrophysics. We elaborate here, from the earlier paradigm, that the concept of induced gravitational collapse is essential to explain the GRB-SN connection. The specific case of a close (orbital period $<1$ h) binary system composed of an evolved star with a Neutron Star (NS) companion is considered. We evaluate the accretion rate onto the NS of the material expelled from the explosion of the core progenitor as a type Ib/c SN, and give the explicit expression of the accreted mass as a function of the nature of the components and binary parameters. We show that the NS can reach, in a few seconds, the critical mass and consequently gravitationally collapses to a Black Hole. This gravitational collapse process leads to the emission of the GRB.
\end{abstract}

\keywords{Type Ib/c Supernovae --- Neutron Star Accretion --- Induced Gravitational Collapse --- Gamma Ray Bursts}

\maketitle

The systematic and spectroscopic analysis of the Gamma Ray Burst (GRB)-Supernova (SN) events, following the pioneering discovery of the temporal coincidence of GRB 980425 \citep{2000ApJ...536..778P} and SN 1998bw \citep{1998Natur.395..670G}, has evidenced the association of other nearby GRBs with Type Ib/c SNe \citep[see e.g.][for a recent review on GRB-SN systems]{2011IJMPD..20.1745D,2011arXiv1104.2274H}. It has been also clearly understood that the most likely explanation of SN Ib/c, which lack Hydrogen (H)/Helium (He) in their spectra, is that the SN core progenitor star, 
likely a He, CO, or a Wolf-Rayet star,
is in a binary system with a compact companion, a Neutron Star (NS) (see e.g.~\cite{1988PhR...163...13N,1994ApJ...437L.115I} and \cite{2007PASP..119.1211F,2010ApJ...725..940Y} for more recent calculations).

In the current literature there has been the attempt to explain both the SN and the GRB as two aspects of the same astrophysical phenomenon. Hence, GRBs have been assumed to originate from a specially violent SN process, a hypernova or a collapsar \citep[see e.g.][and references therein]{2006ARA&A..44..507W}. Both of these possibilities imply a very dense and strong wind-like CircumBurst Medium (CBM) structure. Such a dense medium appears to be in contrast with the CBM density found in most GRBs (see e.g.~Fig.~10 in \cite{2012A&A...543A..10I}). In fact, the average CBM density, inferred from the analysis of the afterglow, has been shown to be in most of the cases of the order of 1 particle cm$^{-3}$ \citep[see e.g.][]{2011IJMPD..20.1797R}. The only significant contribution to the baryonic matter component in the GRB process is the one represented by the baryon load \citep{2000A&A...359..855R}. In a GRB, the electron-positron plasma, loaded with a certain amount of baryonic matter, is expected to expand at ultra-relativistic velocities with Lorentz factors $\Gamma\gtrsim 100$ \citep{1990ApJ...365L..55S,1993MNRAS.263..861P,1993ApJ...415..181M}. Such an ultra-relativistic expansion can actually occur if the amount of baryonic matter, quantifiable through the baryon load parameter, $B=M_B c^2/E_{e^+e-}$, where $M_B$ is the engulfed baryon mass from the progenitor remnant and $E_{e^+e-}$ is the total energy of the $e^+e-$ plasma, does not exceed the critical value $B \sim 10^{-2}$ \citep[see][for details]{2000A&A...359..855R}. 

In our approach we have assumed that the GRB consistently has to originate from the gravitational collapse to a Black Hole (BH). The SN follows instead the complex pattern of the final evolution of a massive star, possibly leading to a NS or to a complete explosion but never to a BH. There is a further general argument in favor of this explanation, namely the extremely different energetics of SNe and GRBs. While the SN energy range is $10^{49}$--$10^{51}$ erg, the GRBs are in a larger and wider range of energies $10^{49}$--$10^{54}$ erg. It is clear that in no way a GRB, being energetically dominant, can originate from the SN. 

There are however scenarios for GRB-SN systems that invoke a single progenitor, e.g.~the collapse of massive stars \citep[see e.g.][for a recent review]{reviewzhang2011}. In these models the core of the star must rotate at very high rates in order to produce, during the gravitational collapse, a collimated (e.g.~jet) emission with a beaming angle $\theta_j$. In this way, an event with observed isotropic energy $E_{iso}$, corresponds to an actual energy released at the source reduced by the beaming factor $f_b =(1-\cos\theta_j)\sim \theta^2_j/2<1$, namely $E_{s}=f_b E_{iso}<E_{iso}$ \citep[see][and references therein]{2006ARA&A..44..507W,reviewzhang2011}. 
Outstandingly small beaming factors of order $f_p\sim 1/500$, corresponding to jet angles $\theta_j \sim 1^\circ$, are needed to bring the most energetics GRBS with $E_{iso}\sim 10^{54}$ erg to standard energies $\sim 10^{51}$ erg \citep{2001ApJ...562L..55F}. However, observational evidence of the existence of so narrow beaming angles in GRBs, as suggested by these models, is inconclusive \citep[see e.g.][]{2006NCimB.121.1171C,2007ApJ...657..359S,2009cfdd.confE..23B}.

We explain the temporal coincidence of the two phenomena, the SN explosion and the GRB, within the concept of \emph{induced gravitational collapse} \citep{2001ApJ...555L.117R,2008mgm..conf..368R}. In the recent years we have outlined two different possible scenarios for the GRB-SN connection. In the first version \citep{2001ApJ...555L.117R}, we have considered the possibility that the GRBs may have caused the trigger of the SN event. For the occurrence of this scenario, the companion star had to be in a very special phase of its thermonuclear evolution \citep[see][for details]{2001ApJ...555L.117R}.

\begin{figure}
\centering
\includegraphics[scale=0.25]{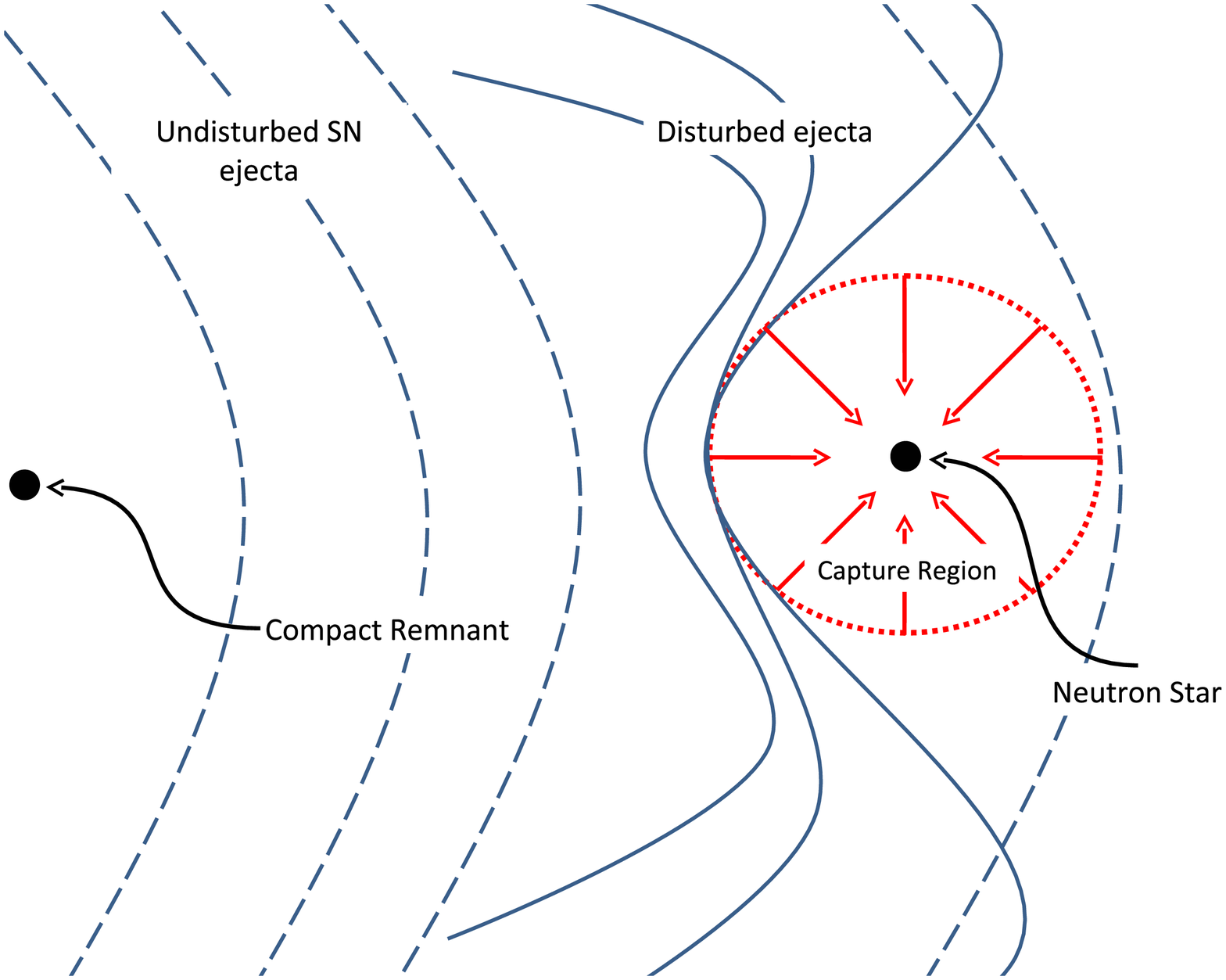}
\caption{Sketch of the accretion induced collapse scenario. An evolved star in close binary with a NS explodes as a SN Ib/c. The NS rapidly accretes a part of the SN ejecta and reaches in a few seconds the critical mass undergoing gravitational collapse to a BH, emitting the GRB.}\label{fig:scenario}
\end{figure}

More recently, we have proposed \citep{2008mgm..conf..368R} a different possibility occurring at the final stages of the evolution of a close binary system: the explosion in such a system of a Ib/c SN leads to an accretion process onto the NS companion. The NS will reach the critical mass value, undergoing gravitational collapse to a BH. The process of gravitational collapse to a BH leads to the emission of the GRB (see Fig.~\ref{fig:scenario}). In this Letter we evaluate the accretion rate onto the NS and give the explicit expression of the accreted mass as a function of the nature of the components and the binary parameters.
%

\begin{figure}
\centering
\includegraphics[scale=0.4]{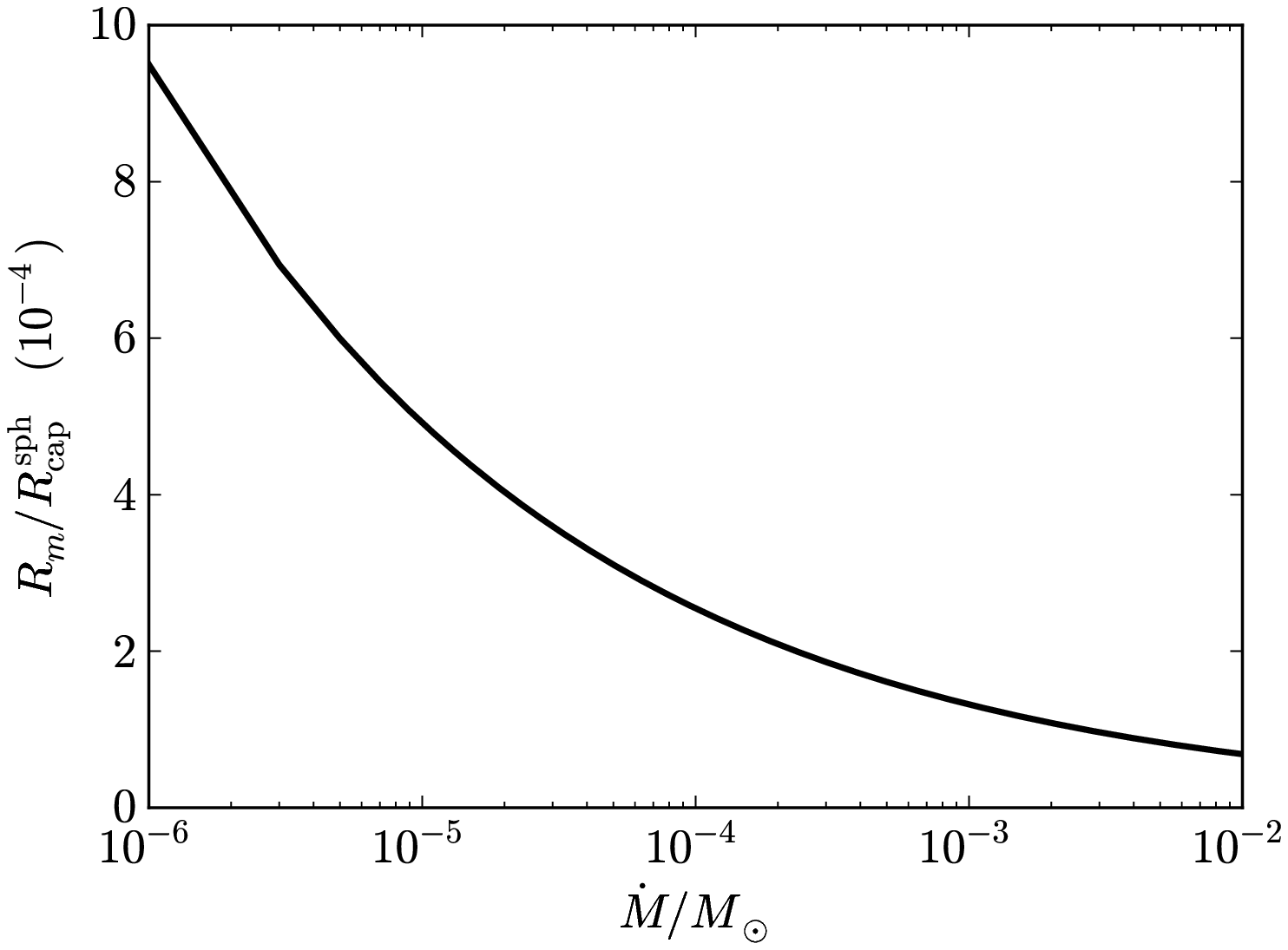}
\caption{Magnetospheric to gravitational capture radius ratio of a NS of $B=10^{12}$ Gauss, $M_{\rm NS}=1.4 M_\odot$, $R=10^6$ cm, in the spherically symmetric case. The flow velocity has been assumed as $v=10^9$ cm s$^{-1}$.}\label{fig:RmRcap}\emph{}
\end{figure}

We turn now to the details of the accretion process of the SN material onto the NS. In a spherically symmetric accretion process, the magnetospheric radius is given by \citep[see e.g.][]{2012MNRAS.420..810T} $R_m = B^2 R^6/(\dot{M} \sqrt{2 G M_{\rm NS}})^{2/7}$, where $B$, $M_{\rm NS}$, $R$ are the NS magnetic field, mass, radius, and $\dot{M}\equiv dM/dt$ is the mass-accretion rate onto the NS. We now estimate the relative importance of the NS magnetic field on the accretion process. At the beginning of a SN explosion, the ejecta moves at high velocities $v\sim 10^9$ cm s$^{-1}$, and the NS will capture matter at a radius approximately given by $R^{\rm sph}_{\rm cap} \sim 2 G M/v^2$. For $R_m << R^{\rm sph}_{\rm cap}$, we can neglect the effects of the magnetic field. In Fig.~\ref{fig:RmRcap} we have plotted the ratio between the magnetospheric radius and the gravitational capture radius as a function of the mass accretion rate onto a NS of $B=10^{12}$ Gauss, $M_{\rm NS}=1.4 M_\odot$, $R=10^6$ cm, and for a flow with velocity $v=10^9$ cm s$^{-1}$. It can be seen how for high accretion rates the influence of the magnetosphere is negligible.

We therefore assume hereafter, for simplicity, that the NS is non-rotating and neglect the effects of the magnetosphere. The NS captures the material ejected from the core collapse of the companion star in a region delimited by the radius $R_{\rm cap}$ from the NS center
\begin{equation}\label{eq:Rcap}
R_{\rm cap} = \frac{2 G M_{\rm NS}}{v^2_{\rm rel,ej}}\, ,
\end{equation}
where $v_{\rm rel,ej}$ is the ejecta velocity relative to the orbital motion of the NS
\begin{equation}\label{eq:vrel}
v_{\rm rel,ej}=\sqrt{v_{\rm orb}^2+v^2_{\rm ej}}\, ,\quad v_{\rm orb}= \sqrt{\frac{G (M_{\rm SN-prog}+M_{\rm NS})}{a}}\, ,
\end{equation}
with $v_{\rm ej}$ the velocity of the ejecta and $v_{\rm orb}$ the orbital velocity of the NS, where $a$ is the binary separation. Here we have assumed that the velocity of the SN ejecta $v_{\rm ej}$ is much larger than the sound speed $c_s$ of the material, namely that the Mach number of the SN ejecta satisfies ${\cal M}=v_{\rm ej}/c_s>>$ 1, which is a reasonable approximation in the present case. The orbital period of the binary system is
\begin{equation}\label{eq:period}
P=\sqrt{\frac{4\pi^2 a^3}{G (M_{\rm SN-prog}+M_{\rm NS})}}\, ,
\end{equation}
where $M_{\rm SN-prog}$ is the mass of the SN core progenitor.

The NS accretes the material that enters into its capture region defined by Eq.~(\ref{eq:Rcap}). The mass-accretion rate is given by \citep[see][for details]{1944MNRAS.104..273B}
\begin{equation}\label{eq:Mdot}
\dot{M}= \xi \pi \rho_{\rm ej} R^2_{\rm cap} v_{ej} = \xi \pi \rho_{ej} \frac{(2 G M_{\rm NS})^2}{(v_{\rm orb}^2+v^2_{ej})^{3/2}}\, , 
\end{equation}
where the parameter $\xi$ is comprised in the range $1/2\leq\xi\leq 1$, $\rho_{\rm ej}$ is the density of the accreted material, and in the last equality we have used Eqs.~(\ref{eq:Rcap}) and (\ref{eq:vrel}). The upper value $\xi=1$ corresponds to the Hoyle-Lyttleton accretion rate \citep{1939PCPS...35..405H}. The actual value of $\xi$ depends on the properties of the medium on which the accretion process occurs, e.g. vacuum, wind. In Fig.~\ref{fig:scenario} we have sketched the accreting process of the SN ejected material onto the NS.

The density of the ejected material can be assumed to decrease in time following the simple power-law \citep[see e.g.][]{1989ApJ...346..847C}
\begin{equation}\label{eq:rhoej}
\rho_{\rm ej}(t)=\frac{3 M_{\rm ej}(t)}{4\pi r^3_{\rm ej}(t)}=\frac{3 M_{\rm ej}}{4\pi \sigma^3 t^{3 n}}\, ,
\end{equation}
where, without loss of generality, we have assumed that the radius of the SN ejecta expands as $r_{\rm ej}=\sigma t^n$, with $\sigma$ and $n$ constants. The velocity of the ejecta is thus $v_{\rm ej}=n r_{\rm ej}/t$.

If the accreted mass onto the NS is much smaller than the initial mass of the ejecta, i.e. $M_{acc}(t)/M_{\rm ej}(0)<<1$, one can assume $M_{\rm ej}(t)\approx M_{\rm ej}(0)$ and thus the integration of Eq.~(\ref{eq:Mdot}) gives

\begin{equation}\label{eq:deltaM}
\Delta M (t)=\left.\int_{t^{\rm acc}_0}^t \dot{M} dt = \pi \xi (2 G M_{\rm NS})^2\frac{3 M_{\rm ej}(0)}{4\pi n^3 \sigma^6} {\cal F}\right|_{t^{\rm acc}_0}^t\, ,
\end{equation}
where 
\begin{widetext}
\begin{equation}
{\cal F} = \frac{t^{-3 (n+1)} \left[-4 n (2 n-1) t^{4 n} \sqrt{k t^{2-2 n}+1} \, _2F_1\left(\frac{1}{2},\frac{1}{n-1};\frac{n}{n-1};-k
   t^{2-2 n}\right)-k^2 \left(n^2-1\right] t^4+2 k (n-1) (2 n-1) t^{2 n+2}+4 n (2 n-1) t^{4 n}\right)}{k^3 (n-1) (n+1) (3 n-1)\sqrt{k+t^{2 n-2}}}\, ,
\end{equation}
\end{widetext}
with $k=v^2_{\rm orb}/(n\,\sigma)^2$, $_{2}F_{1}(a,b;c;z)$ is the Hypergeometric function, and $t^{\rm acc}_0$ is the time at which the accretion process starts, namely the time at which the SN ejecta reaches the NS capture region (see Fig.~\ref{fig:scenario}), i.e. so $\Delta M (t)=0$ for $t \leq t^{\rm acc}_0$. The above expression increases its accuracy for massive NSs close to the critical value, since the amount of mass needed to reach the critical mass by accretion is much smaller than $M_{\rm ej}$. In general, the total accreted mass must be computed from the full numerical integration of Eq.~(\ref{eq:Mdot}).

We turn now to the maximum stable mass of a NS. Non-rotating NS equilibrium configurations have been recently constructed by \cite{belvedere2012} taking into account the strong, weak, electromagnetic, and gravitational interactions within general relativity. The equilibrium equations are given by the general relativistic Thomas-Fermi equations, coupled with the Einstein-Maxwell system of equations, the Einstein-Maxwell-Thomas-Fermi system of equations, which must be solved under the condition of global charge neutrality. 
The strong interactions between nucleons have been modeled through the exchange of virtual mesons ($\sigma$, $\omega$, $\rho$) within the Relativistic Mean Field (RMF) model, in the version of \citet{1977NuPhA.292..413B}. These self-consistent equations supersede the traditional Tolman-Oppenheimer-Volkoff ones that impose the condition of local charge neutrality throughout the configuration. 

\begin{figure}
\centering
\includegraphics[scale=0.4]{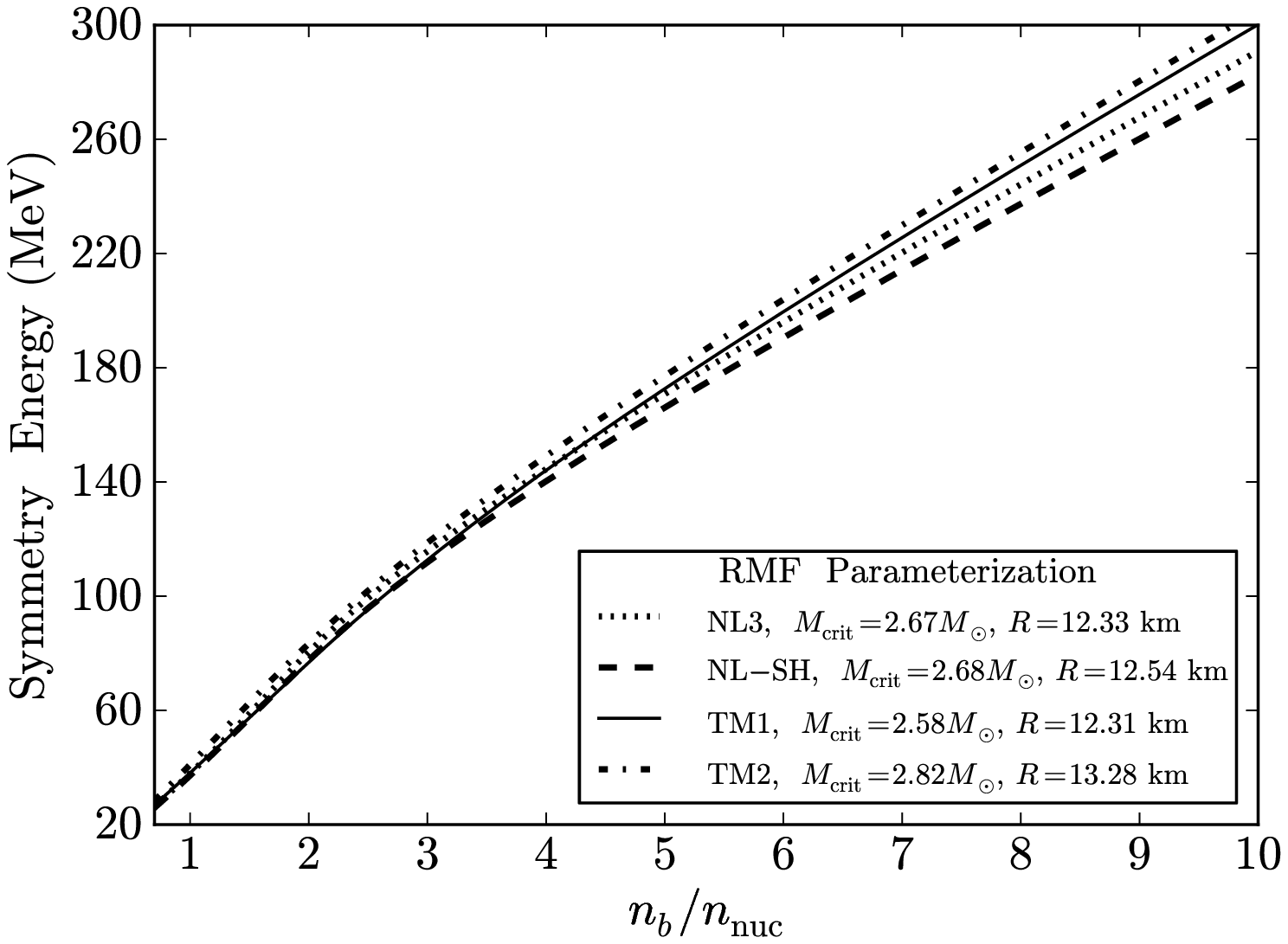}
\caption{Nuclear symmetry energy as a function of the baryon density for selected parameterizations of the RMF nuclear model \citep[see][for details]{belvedere2012}.}\label{fig:esymrho}
\end{figure}

The uncertainties in the behavior of the nuclear equation of state (EOS) at densities about and larger than the nuclear saturation density $n_{\rm nuc}\approx 0.16$ fm$^{-3}$, lead to a variety of EOS with different nuclear parameters. A crucial parameter in this respect is the so-called nuclear symmetry energy, $\left.E_{\rm sym}=d^2 ({\cal E}/n_b)/d\delta^2\right|_{\delta=0,n_{\rm nuc}}$, where ${\cal E}$ is the nuclear matter energy-density, and $\delta=(n_n-n_p)/n_b$ is the asymmetry parameter with $n_n$, $n_p$, $n_b=n_p+n_p$ the neutron, proton, baryon densities; we refer to \citep{2012PhRvC..86a5803T} for a recent review. The symmetry energy is relevant for the determination of the value and density dependence of the particle abundances (e.g.~$n_n/n_p$ ratio) in the NS interior \citep[see e.g.][]{1987PhLB..199..469M,2007PhRvC..76b5801K,2009PhLB..682...23S,2010PhRvL.105p1102H,2011PhRvC..83f5809L}. The differences in the behavior of $E_{\rm sym}(n)$ for different nuclear EOS models and parameterizations lead to a variety of NS mass-radius relations and consequently to different values the NS critical mass $M_{\rm crit}$ and the corresponding radii \citep[see e.g.][]{2012PhRvC..85c2801G}. We have plotted in Fig.~\ref{fig:esymrho} the behavior nuclear symmetry energy for the RMF parameterizations used in \citep{belvedere2012} in a wide range of baryon densities expanding from $n_b\sim 0.7 n_{\rm nuc}$ all the way up to high densities $n_b\sim 10 n_{\rm nuc}$ found in the cores of NSs; we have included in the legend the values of the mass and radius of the critical neutron star configuration, $M_{\rm crit}$ and $R$. Concerning our induced gravitational collapse scenario, the precise value of the time needed for the NS to reach $M_{\rm crit}$ by accretion of the SN material depends, for fixed binary parameters $(M_{\rm NS}, a, M_{\rm prog}, v_{\rm ej})$, on the adopted EOS which lead to a specific value of $M_{\rm crit}$.

The high and rapid accretion rate of the SN material can lead the NS mass to reach the critical value $M_{\rm crit}$. This system will undergo gravitational collapse to a BH, producing a GRB. The initial NS mass is likely to be rather high due to the highly non-conservative mass transfer during the previous history of the evolution of the binary system \citep[see e.g.][for details]{1988PhR...163...13N,1994ApJ...437L.115I}. Thus, the NS could reach the critical mass in just a few seconds. Indeed, Eq.~(\ref{eq:Mdot}) shows that for an ejecta density $10^6$ g cm$^{-3}$ and ejecta velocity $10^9$ cm s$^{-1}$, the accretion rate might be as large as $\dot{M} \sim 0.1 M_\odot s^{-1}$.

The occurrence of a GRB-SN event in the scenario presented in this Letter is subjected to some specific conditions of the binary progenitor system, such as a short binary separation and orbital period $P<1$ h. This is indeed the case of GRB 090618 \citep{2012A&A...543A..10I} and GRB 970828 \citep{2012arXiv1205.6651I}, which we are going to analyze within the framework presented here in forthcoming publications. In addition of offering an explanation to the GRB-SN temporal coincidence, the considerations presented in this Letter leads to an astrophysical implementation of the concept of proto-BH, generically introduced in our previous works on of GRBs 090618, 970828, and 101023 \citep[see][respectively]{2012A&A...543A..10I,2012arXiv1205.6651I,2012A&A...538A..58P}. The proto-BH represents the first stages, $20 \lesssim t \lesssim 200$ s, of the SN evolution.

It is also worth noticing that the condition $B \lesssim 10^{-2}$ on the baryon load parameter of a GRB \citep{2000A&A...359..855R} might constrain on the binary separation $a$ for the occurrence a GRB-SN event. When the NS reaches the critical mass, the distance between the location of the front of the undisturbed SN ejecta and the NS center should be $<< a$, otherwise the emitted $e^+e^-$ plasma in the GRB might engulf a large amount of baryonic matter from the SN ejecta, reaching or even overcoming the critical value $B \sim 10^{-2}$.

It is appropriate now to discuss the possible progenitors of such binary systems. A viable progenitor is represented by X-Ray Binaries such as Cen X-3 and Her X-1 \citep{1972ApJ...172L..79S,1972ApJ...174L..27W,1972ApJ...174L.143T,1973ApJ...180L..15L,1975ASSL...48.....G,2011ApJ...730...25R}. The binary system is expected to follow an evolutionary track \citep[see][for details]{1988PhR...163...13N,1994ApJ...437L.115I}: the initial binary system is composed of main-sequence stars 1 and 2 with a mass ratio $M_2/M_1\gtrsim 0.4$. The initial mass of the star 1 is likely $M_1 \gtrsim 11 M_\odot$, leaving a NS through a core-collapse event. The star 2, now with $M_2\gtrsim 11 M_\odot$ after some almost conservative mass transfer, evolves filling its Roche lobe. It then starts a spiral in of the NS into the envelope of the star 2. If the binary system does not merge, it will be composed of a Helium star and a NS in close orbit. The Helium star expands filling its Roche Lobe and a non-conservative mass transfer to the NS, takes place. This scenario naturally leads to a binary system composed of a CO star and a massive NS, as the one considered in this Letter.

We point out that the systems presenting a temporal coincidence of GRB-SN form a special class of GRBs: 

(1) There exist Ib/c SNe not associated to a GRB, e.g.~the observations of SN 1994I \citep{2002ApJ...573L..27I} and SN 2002ap \citep{2004A&A...413..107S}. Also this class of apparently isolated SNe may be in a binary system with a NS companion at a large binary separation $a$ and long orbital period $P$ (\ref{eq:period}), and therefore the accretion rate (\ref{eq:Mdot}) is not highly enough to trigger the process of gravitational collapse of the NS. 
A new NS binary system may be then formed and lead the emission of a short GRB in a NS merger after the shrinking of the binary orbit by gravitational waves emission.

(2) There are GRBs that do not show the presence of an associated SN. This is certainly the case of GRBs at large cosmological distances $z\gtrsim 0.6$ when the SN is not detectable even by the current high power optical telescopes. This is likely the case of GRB 101023 \citep{2012A&A...538A..58P}. 

(3) There is the most interesting case of GRBs that do not show a SN, although it would be detectable. This is the case of GRB 060614 \citep{2009A&A...498..501C} in which a possible progenitor has been indicated in a binary system formed of a white dwarf and a NS, which clearly departs from the binary class considered in this Letter. Finally, there are systems originating genuinely short GRBs which have been proved to have their progenitors in binary NSs, and clearly do not have an associated SN, e.g.~GRB 090227B \citep{2012arXiv1205.6600M,2012arXiv1205.6915R}.

Before closing, we like to look to the problem of the remnants of the class of GRBs considered in this Letter. It is clear that after the occurrence of the SN and the GRB emission, the outcome is represented, respectively, by a NS and a BH. A possible strong evidence of the NS formation is represented by the observation of a characteristic late ($t=10^8$--$10^9$ s) X-ray emission (called URCA sources, see \cite{2005tmgm.meet..369R}) that has been interpreted as originated by the young ($t \sim$ 1 minute--$(10$--$100)$ years), hot ($T \sim 10^7$--$10^8$ K) NS, which we have called neo-NS \citep[see][for details]{2012A&A...540A..12N}. This has been indeed observed in GRB 090618 \citep{2012A&A...543A..10I} and also in GRB 101023 \citep{2012A&A...538A..58P}. If the NS and the BH are gravitationally bound they give origin to a new kind of binary system, which can lead itself to the merging of the NS and the BH and consequently to a new process of gravitational collapse of the NS into the BH. In this case the system could originate a yet additional process of GRB emission and possibly a predominant emission in gravitational waves.

We thank the anonymous referee for many suggestions which have improved the presentation of our results.

\end{document}